\documentclass[10pt]{IEEEtran}

\usepackage{graphicx} 
\usepackage{amsmath,amssymb}
\usepackage{amsfonts}
\usepackage{mathrsfs}
\usepackage[noadjust]{cite}
\usepackage{xcolor}
\usepackage{comment}
\usepackage{enumitem}

\usepackage{color}

\newtheorem{theorem}{Theorem}
\newtheorem{lemma}[theorem]{Lemma}
\newtheorem{claim}[theorem]{Claim}

\newtheorem{corollary}[theorem]{Corollary}
 
\newtheorem{example}{Example} 
\newtheorem{remark}{Remark}

\allowdisplaybreaks

\newcommand{\be}{\begin{equation}}
\newcommand{\ee}{\end{equation}}
\newcommand{\calA}{\mathcal{A}}
\newcommand{\calE}{\mathcal{E}}
\newcommand{\calU}{\mathcal{U}}
\newcommand{\calV}{\mathcal{V}}
\newcommand{\calX}{\mathcal{X}}
\newcommand{\calY}{\mathcal{Y}}
\newcommand{\hatM}{\hat{M}}
\newcommand{\hatm}{\hat{m}}
\newcommand{\hatX}{\hat{X}}
\newcommand{\hatx}{\hat{x}}
\newcommand{\eps}{\epsilon}
\newcommand{\bbR}{\mathbb{R}}

\newcommand{\cX}{{\cal X}}
\newcommand{\cY}{{\cal Y}}

\title{On the Benefit of Cooperation in Relay Networks}

\author{Oliver Kosut, Michelle Effros, Michael Langberg
\thanks{O. Kosut is with the School of Electrical, Computer and Energy Engineering at Arizona State University. Email: {\tt okosut@asu.edu}}
\thanks{M. Effros is with the Department of Electrical Engineering at the California Institute of Technology.
Email: \texttt{effros@caltech.edu}}
\thanks{M. Langberg is with the Department of Electrical Engineering at the University at Buffalo (State University of New York).  
Email: \texttt{mikel@buffalo.edu}}
\thanks{This work is supported in part by NSF grants CCF-1817241, CCF-1908725, and CCF-1909451. 
}
}

\begin{document}

\maketitle

\begin{abstract}
This work addresses the cooperation facilitator (CF) model, in which network nodes coordinate through a rate limited communication device. For independent multiple-access channel (MAC) encoders, the CF model is known to show significant rate benefits, even when the rate of cooperation is negligible.
Specifically, the benefit in MAC sum-rate, as a function of the cooperation rate $C_{CF}$, sometimes has an infinite slope at $C_{CF}=0$.
This work studies the question of whether cooperation through a CF can yield similar infinite-slope benefits when applied to internal network encoders in which {\em dependence} among MAC transmitters can be established without the help of the CF.
Towards this end, this work studies the CF model when applied to relay nodes of a single-source, single-terminal, diamond network consisting of a broadcast channel followed by a MAC.
In the relay channel with orthogonal receiver components, careful generalization of the partial-decode-forward/compress-forward lower bound to the CF model yields sufficient conditions for an infinite-slope benefit.
Additional results include derivation of a family of diamond networks for which the infinite-slope rate-benefit derives directly from the properties of the corresponding MAC component when studied in isolation.
\end{abstract}

\section{Introduction}

The information theory and communication literatures
approach the goal of improving network communication performance
in a variety of ways.
While some studies investigate
how to get the best possible performance out of existing networks,
others seek better designs for future networks.
In practice, the way that networks improve over time
is somewhere in between ---
a combination of adding new resources
and making better use of what is already there.
We here seek new tools for guiding that process,
focusing on the questions of whether and where
small changes to an existing network
can have a big impact on network capacity.

One example of a network in which incremental network modifications
can achieve radical network improvement,
introduced in~\cite{NEL:18},
employs the multiple-access channel (MAC)
and a node called a cooperation facilitator (CF).
In practice, the CF is any communicating device
that can receive information from multiple transmitters.
In any MAC for which dependent channel inputs from the transmitters
would yield a higher mutual information
between the MAC's inputs and output
than is achievable with the independent channel inputs
employed in calculating the MAC capacity,
adding a small communication link from the CF
to either or both of the transmitters yields
a disproportionately large capacity improvement~\cite{NEL:18}.
Specifically, the curve describing
the improvement in MAC sum-capacity
as a function of the capacity $C_{CF}$
of the cooperation-enabling CF output link
has slope infinity at $C_{CF}=0$~\cite[Theorem~3]{NEL:18}.
In some cases, even a single bit ---
not rate~1, but a single bit no matter what the blocklength,
suffices to change the network capacity~\cite{langberg2016capacity,KEF:21}.

Since the infinite-slope improvement in the MAC-capacity
results from creating dependence
where none could otherwise be observed,
it is tempting to believe that the infinite-slope phenomenon
cannot occur either in cases
where dependence is already attainable
or where dependence is not critical
to attaining the best possible performance.
In this paper, we explore these two intuitions ---
seeking to understand whether incremental changes
can achieve disproportionate channel benefits
in these scenarios.

Toward this end, we investigate a single coding framework
where both scenarios can arise.
We pose this framework as a diamond network
in which a single transmitter communicates to a collection of relays,
and the relays work independently to transmit information
to a shared receiver.
Since the communication goal in the diamond network
is to transmit information
from a single transmitter
at the start of the diamond network
to a single receiver at its end,
dependence at the relays may be available naturally;
we investigate whether this availability
precludes the possibility of incremental change
with disproportionate impact.
When the links from relays to the receiver
are independent, point-to-point channels,
the resulting degenerate MAC
fails to meet the prior condition
specifying that input dependence should increase sum capacity;
we  investigate whether this failure precludes
the desired small cost, large benefit tradeoff
to incremental network modifications.

\begin{figure}
\begin{center}
\includegraphics[width=\columnwidth]{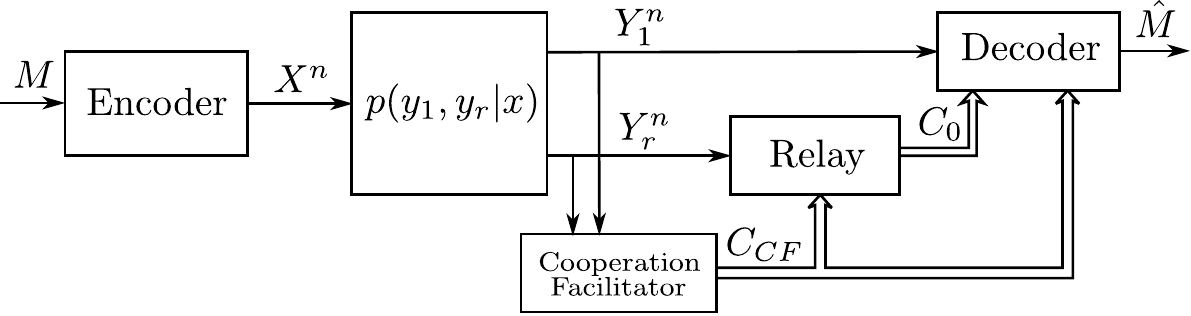}
\end{center}
\caption{Relay channel with orthogonal receiver components and a cooperation facilitator, a special case of the general diamond relay network.}
\label{fig:relay_channel}
\end{figure}

The rest of this paper is organized as follows. {In Section~\ref{sec:problem}, we set up the problem of the diamond relay network with $N$ relay nodes and a cooperation facilitator (Fig.~\ref{fig:diamond}), which allows us to pose our main question about the power of cooperation in a relay network. Our results focus on two special cases of this network. The first, covered in Section~\ref{sec:relay}, is the relay channel with orthogonal receiver components (Fig.~\ref{fig:relay_channel}). Here, we present an achievability bound for the CF problem, as well as sufficient conditions for the infinite-slope phenomenon to occur. In Section~\ref{sec:3relay}, we explore a 3-relay example (Fig.~\ref{fig:diamond1}) that allows us to exploit the results of \cite{NEL:18} on the MAC to demonstrate the infinite-slope phenomenon in a larger network with only one source.
}

\section{Problem Setup}\label{sec:problem}

\begin{figure*}
\begin{center}
\includegraphics[width=.7\textwidth]{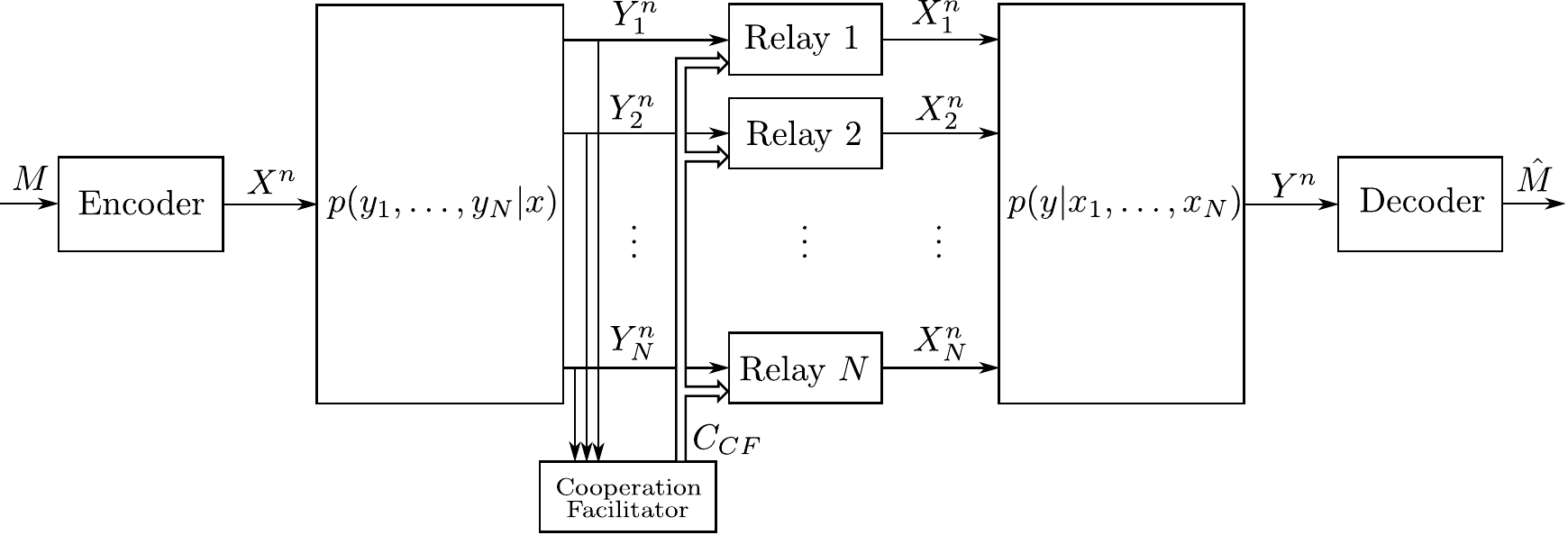}
\end{center}
\caption{General diamond relay network with $N$ nodes and a cooperation facilitator (CF).}
\label{fig:diamond}
\end{figure*}

\emph{Notation}: For any integer $k$, $[k]$ denotes the set $\{1,2,\ldots,k\}$. Capital letters (e.g., $X$) denote random variables, lower-case letters (e.g., $x$) denote realizations of the corresponding variable, and calligraphic letters (e.g., $\calX$) denote the corresponding alphabet. Vectors are denoted with superscript (e.g., $x^n=(x_1,\ldots,x_n)$). We use standard notation for mutual information and entropy.

A diamond relay network with $N$ relay nodes and a cooperation facilitator (CF)---shown in Fig.~\ref{fig:diamond}---is given by a broadcast channel $p(y_1,\ldots,y_N|x)$, followed by a MAC $p(y|x_1,\ldots,x_N)$. An $(n,R)$ code for the diamond relay network is composed of
\begin{itemize}[leftmargin=12pt,itemsep=2pt]
\item an encoder $f:[2^{nR}]\to\calX^n$,
\item a CF-encoder $f_{CF}:\prod_{j\in[N]} \calY_j^n\to [2^{nC_{CF}}]$,
\item a relay encoder $f_j:\calY_j^n\times [2^{nC_{CF}}]\to \calX_j^n$ for each $j\in [N]$,
\item a decoder $g:\calY^n\to [2^{nR}]$.
\end{itemize}
The message $M$ is assumed to be uniformly drawn from $[2^{nR}]$. Encoded message $X^n=f(M)$ is transmitted by the encoder into the broadcast channel, which outputs $Y_j^n$ at relay $j$, $j\in[N]$. The CF observes all the outputs of the broadcast channel, and encodes $K=f_{CF}(Y_1^n,\ldots,Y_N^n)$, which is sent to each relay. Relay $j$ encodes $X_j^n=f_j(Y_j^n,K)$ and transmits it into the MAC. Finally, the output signal $Y^n$ is received and decoded to $\hatM=g(Y^n)$. The overall probability of error is given by $P_e^{(n)}=P(M\ne\hatM)$. We say a rate $R$ is \emph{achievable} if there exists a sequence of $(n,R)$ codes with $P_e^{(n)}\to 0$. The \emph{capacity} $C(C_{CF})$ is the supremum of all achievable rates for a given CF capacity $C_{CF}$. This function is non-decreasing in $C_{CF}$, and so its derivative $C'(C_{CF})$ is non-negative.

We are interested in characterizing $C(C_{CF})$, but more specifically, we are focused on the following question:

\emph{Main question:} For a given network, is $C'(0)=\infty$?

\section{Relay Channel with Orthogonal Receiver Components}\label{sec:relay}

The first special case of the diamond relay network that we focus on is the relay channel with orthogonal receiver components. Here, we specialize the general model described above in several ways. First, we assume there are only $N=2$ relay nodes, and further we assume that the received signal at the decoder is made up of orthogonal components, one from each relay. That is, $Y=(Y_a,Y_b)$, where the MAC model factors as
\be
p(y_a,y_b|x_1,x_2)=p(y_a|x_1) p(y_b|x_2).
\ee
Given this factorization, the capacity of the overall network depends on the channels from $X_1$ to $Y_a$ and from $X_2$ to $Y_b$ only through their capacities~\cite{KoetterE:11}. 
Thus, we can simplify the problem by replacing these noisy channels with rate-limited bit-pipes of capacities $C_1$ and $C_0$ from each relay to the decoder. Finally, we assume that $C_1=\infty$; i.e., we assume that Relay 1 is able to transmit all of its information (consisting of $Y_1^n$ as well as the CF signal $K$) directly to the decoder. These simplifications yield the network model shown in Fig.~\ref{fig:relay_channel}.  Note that this network only has one relay node, so it makes sense to call it a relay channel model rather than a diamond network model. We have also relabelled $Y_2$ as $Y_r$ to emphasize that it is the relay's received signal; this also makes the notation consistent with \cite{ElGamal2021,ElGamal2021a}.

\subsection{Main Achievability Result}
\label{sec:2relay}

\begin{theorem}\label{thm:achievability}
Consider a relay channel with orthogonal receiver components with broadcast channel distribution $p(y_r,y_1|x)$, capacity $C_0$ from relay to destination, and the CF capacity $C_{CF}$. Rate $R$ is achievable if
\begin{align}
R&\le I(U;Y_r)+\min\{I(X;Y_1,Y_r|U),I(X;Y_1,V|U)\},\label{rate_condition1}\\
R&\le\min\{I(U;Y_1),I(U,Y_r)\}+I(X;Y_1|U)\nonumber
\\&\qquad+I(V;X,Y_1|U)-I(Y_r;V|U)+C_0,\label{rate_condition2}\\
C_{CF}&\ge I(X,Y_1;V|U,Y_r),\label{CF_condition}
\end{align}
for some distribution
\be
p(u,x) p(y_r,y_1|x) p(v|x,y_1,y_r,u).
\ee
\end{theorem}

\begin{IEEEproof}
See Appendix~\ref{appendix:achievability}.
\end{IEEEproof}

\begin{remark}\label{remark:pdc}
Thm.~\ref{thm:achievability} reduces to the well-known combined partial-decode-forward/compress-forward lower bound for the standard problem (without a CF), which originated in \cite{Cover1979}. For the relay channel with orthogonal receiver components, \cite{ElGamal2021} showed that this classical bound can be written as follows:  rate $R$ is achievable if
\begin{align}
    R&\le I(U;Y_r)+I(X;Y_1,V|U),\label{pdc_bd1}\\
    R&\le \min\{I(U;Y_1),I(U;Y_r)\}+I(X;Y_1|U)+C_0\nonumber\\ &\qquad-I(Y_r;V|U,X,Y_1)\label{pdc_bd2}
\end{align}
for some distribution
\be\label{pdc_bd_distribution}
p(u,x)p(y_r,y_1|x) p(v|y_r,u).
\ee

In Thm.~\ref{thm:achievability}, removing the CF  is equivalent to setting $C_{CF}=0$. Thus, \eqref{CF_condition} implies the Markov chain $(X,Y_1)-(U,Y_r)-V$, which implies that the variables have a joint distribution that factors as \eqref{pdc_bd_distribution}. Thus, by the data processing inequality, $I(X;Y_1,V|U)\le I(X;Y_1,Y_r|U)$, so \eqref{rate_condition1} becomes \eqref{pdc_bd1}. In addition,
\begin{align}
&I(V;X,Y_1|U)-I(Y_r;V|U)
\\&=-H(V|U,X,Y_1)+H(V|U,Y_r)
\\&=-H(V|U,X,Y_1)+H(V|U,Y_r,X,Y_1)
\\&=-I(Y_r;V|U,X,Y_1),
\end{align}
so \eqref{rate_condition2} becomes \eqref{pdc_bd2}.
\end{remark}

\subsection{Sufficient Conditions for Infinite Slope}
\label{sec:slope}

The following theorem provides a sufficient condition for which, given a starting achievable point for the partial-decode-forward/compress-forward bound without cooperation (i.e., the bound in \eqref{pdc_bd1}--\eqref{pdc_bd2}), the achievable rate from Thm.~\ref{thm:achievability} with cooperation improves over the starting point and this improvement has infinite slope as a function of $C_{CF}$.

\begin{theorem}\label{thm:infinite_slope}
Fix a distribution $p(u,x)p(v|y_r,u)$. Let $R$ be a rate satisfying the no-cooperation achievability conditions in \eqref{pdc_bd1}--\eqref{pdc_bd2} for this distribution. Suppose 
$I(X;Y_1,V|U)<I(X;Y_1,Y_r|U)$, and 
there does \emph{not} exist $\lambda\in[0,1]$ and $\gamma(u,x,y_1,y_r)\in\bbR$ for each $x,y_1,y_r,u$ such that
\begin{gather}
p(v|u,x,y_1)=\frac{p(v|u,y_1)^\lambda p(v|u,y_r)^{1-\lambda}}{\gamma(u,x,y_1,y_r)}\label{infinite_slope_condition}
\end{gather}
for all $u,x,y_1,y_r,v$ where $p(u,x,y_1,y_r)>0,p(v|y_r,u)>0$. Then
\be\label{infinite_slope}
\lim_{C_{CF}\to 0} \frac{C(C_{CF})-R}{C_{CF}} = \infty.
\ee
\end{theorem}

\begin{IEEEproof}
See Appendix~\ref{appendix:infinite_slope}.
\end{IEEEproof}

\begin{remark}
We note that there are two ways for \eqref{infinite_slope} to hold: (1) $C(0)>R$; that is, the rate $R$, while achievable without cooperation, is smaller than the no-cooperation capacity of the relay channel; (2) $C(0)=R$, and $C'(0)=\infty$. Here, the rate $R$ is the no-cooperation capacity, so \eqref{infinite_slope} indicates that the CF really can improve the capacity of the relay network in an infinite-slope manner. Thus, this latter case is the one we are particularly interested in, as it gives an affirmative answer to the Main Question. Unfortunately, for any problem instance for which a matching converse for the no-cooperation setting is unavailable, even if \eqref{infinite_slope} holds, there is no way to know which situation we are in. Still, if $R$ represents the best-known achievable rate for a given network, \eqref{infinite_slope} has a non-trivial consequence, showing that the state-of-the-art can be improved disproportionately by a small amount of cooperation.
\end{remark}

While the condition in Thm.~\ref{thm:infinite_slope} is sometimes hard to verify, the following corollary provides a simpler sufficient condition for the same conclusion.

\begin{corollary}\label{corollary:all_positive_channel}
Assume that $p(y_1,y_r|x)>0$ for all letters $x,y_1,y_r$. Consider any distribution $p(u,x)p(v|u,y_r)$. Let $R$ be a rate satisfying \eqref{pdc_bd1}--\eqref{pdc_bd2} for this distribution. Then at least one of the following possibilities hold:
\begin{enumerate}
    \item there exists a function $g:\calU\times\calY_r\to\calV$ where rate $R$ satisfies \eqref{pdc_bd1}--\eqref{pdc_bd2} with $V=g(U,Y_r)$.
    \item \eqref{infinite_slope} holds.
\end{enumerate}
$R_{\text{PD/C}}(0)$ $p(v|u,y_r)$ is non-deterministic; i.e., $0<p(v|u,y_r)<1$ for some $u,v,y_r$. Then $R'_{\text{PD/C}}(0)=\infty$.
\end{corollary}

\begin{IEEEproof}
See Appendix~\ref{appendix:all_positive_channel}.
\end{IEEEproof}

\subsection{Example Relay Channels}

For some relay channels, \cite{Aleksic2009} showed that the compress-forward bound achieves capacity. Thus, it is possible to definitively answer the Main Question for these channels. The following example illustrates one such channel.

\begin{example}
Let $X\in\{0,1\}$, $Y_1=X\oplus Z$, $Y_r=Z\oplus W$, where $\oplus$ indicates modulo-2 addition, $Z\sim\text{Ber}(p)$, $W\sim\text{Ber}(\delta)$, and $X,Z,W$ are mutually independent. 
For this channel, the capacity without cooperation is shown in \cite{Aleksic2009} to be given by
\be\label{example_capacity}
C(0)=\max_{p(v|y_r):I(Y_r;V)\le C_0} 1-H(Z|V).
\ee
Moreover, this rate is achieved by compress-forward by choosing $X\sim\text{Ber}(1/2)$ and setting $p(v|y_r)$  to be the distribution achieving the maximum in \eqref{example_capacity}.  Corollary~\ref{corollary:all_positive_channel} applies to this channel, since $p(y_1,y_r|x)>0$ for all $(x,y_1,y_r)$ as long as $p,\delta\in(0,1)$. Moreover, the only deterministic distributions from $Y_r$ to $V$ are where either $V$ is a constant, or $V=Y_r$ (or equivalent). It is easy to see that as long as $0<C_0<H(Y_r)=H(p\oplus\delta)$, neither of these choices for $V$ is optimal. Therefore, in all non-trivial cases, $C'(0)=\infty$ for this channel.
\end{example}

The following relay channel example is one for which the no-cooperation capacity is not known. However, we can verify the sufficient condition from Thm.~\ref{thm:infinite_slope}, thus showing that an infinite-slope improvement is possible through cooperation.

\begin{example}
\label{example:erasure}
Let $X\in\{0,1\}$, and let $p(y_1,y_r|x)=p(y_1|x)p(y_r|x)$, where each of the two component channels is a binary erasure channel (BEC) with erasure probability $p$. An achievable rate for the no-cooperation case from \eqref{pdc_bd1}--\eqref{pdc_bd2} is given by taking $U=\emptyset$, $X$ to be uniform on $\{0,1\}$, and $p(v|y_r)$ to be a channel that further erases any un-erased bit with probability $q$; that is,
\be
p(v|y_r)=\begin{cases} 1-q & v=y_r\in\{0,1\}\\
q & v=e,\ y_r\in\{0,1\},\\
1 & v=y_r=e,\\
0 & \text{otherwise}.\end{cases}
\ee
This leads to the achievable rate
\begin{multline}
R=\max_{q\in[0,1]} \min\{(1-p)(1+p(1-q)),
\\1-p-H((1-p)(1-q))+(1-p)H(q)+C_0\}
\end{multline}
where $H(\cdot)$ is the binary entropy function.

Note that this channel does not satisfy the conditions of Corollary~\ref{corollary:all_positive_channel}, since $p(y_1,y_r|x)$ is not always positive. Instead, we verify the sufficient condition of Thm.~\ref{thm:infinite_slope} directly. Suppose that there exists a $\lambda$ and $\gamma$ satisfying \eqref{infinite_slope_condition}. Note that $Y_1-X-Y_r-V$ is a Markov chain, so $p(v|x,y_1)=p(v|x)$. For $x\in\{0,1\}$ and any $y_1\in\{x,e\}$
\begin{align}
&\frac{p_{V|X}(x|x)}{p_{V|X}(e|x)}
=\frac{(1-p)(1-q)}{1-(1-p)(1-q)}\label{bec_condition_lhs}
\\&=\frac{p_{V|Y_1}(x|y_1)^\lambda (1-q)^{1-\lambda}}{\gamma'(x,y_1,x)}\cdot \frac{\gamma'(x,y_1,x)}{p_{V|Y_1}(e|y_1)^\lambda q^{1-\lambda}}
\\&=\left(\frac{1-q}{q}\right)^{1-\lambda} \left(\frac{p_{V|Y_1}(x|y_1)}{p_{V|Y_1}(e|y_1)}\right)^\lambda
\\&=\left(\frac{1-q}{q}\right)^{1-\lambda}
\begin{cases} 
\left(\frac{(1-p)(1-q)}{1-(1-p)(1-q)}\right)^\lambda, & y_1=x\\
\left(\frac{\frac{1}{2}(1-p)(1-q)}{1-(1-p)(1-q)}\right)^\lambda, & y_1=e.\end{cases}
\label{bec_condition}
\end{align}
This cannot hold with equality for both $y_1=x$ and $y_1=e$ unless $p=0$, $p=1$, or $q=1$. Therefore, except in these trivial cases, infinite slope improvement occurs.
\end{example}

\section{3-relay network example}\label{sec:3relay}

In the analysis of Theorem~\ref{thm:achievability} and $C'(0)$ for the orthogonal-receiver setting given in Sections~\ref{sec:2relay} and \ref{sec:slope},  relay-cooperation is governed by the statistics of the broadcast channel $p(y_1,y_r|x)$ and, roughly speaking, is designed to ``remove'' from  $Y_r^n$ message information that can be obtained at the receiver from $Y_1^n$. 
In this aspect, we say that the design of cooperation information looks {\em backwards} and is governed by the broadcast channel of the diamond network. 

\begin{figure}
\includegraphics[width=1.8\columnwidth]{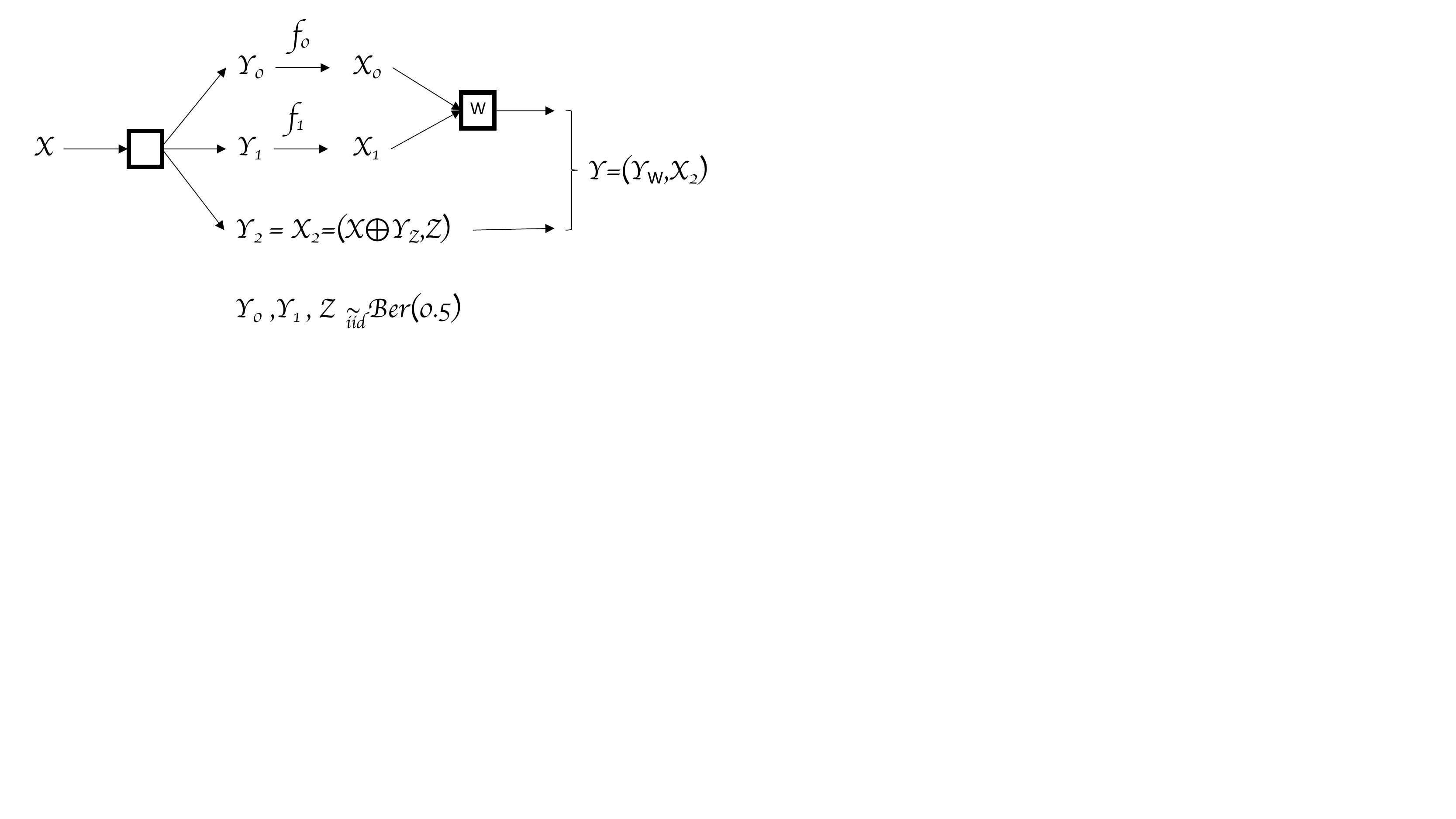}
\vspace{-5.5cm}
\caption{The 3-relay diamond network of Claim~\ref{claim:3relay}.}
\label{fig:diamond1}
\end{figure}

In this section, we study a {\em forward} form of cooperation, that takes into account the MAC appearing in the second stage of the diamond network.
For forward-looking cooperation, it is tempting to treat the MAC ``in isolation'', rather than as part of a larger network, and to design cooperation solely based on the MAC noise statistics, as done in \cite{NEL:18,KEF:21}.
In general, designing cooperation by treating the MAC as an isolated component  
may not suffice to improve communication of the diamond network, because  MAC encoders in the diamond network potentially hold dependent information resulting from the broadcast stage of communication.
Nevertheless, in what follows, we present a family of 3-relay diamond networks for which  cooperation-gain in the network as a whole is derived directly from the MAC cooperation-gain when studied in isolation; the latter is  well understood and given in \cite{NEL:18}. 
Our network family is described below and depicted in Figure~\ref{fig:diamond1}.

Consider the {\em diamond network}  defined by broadcast channel $(\cX,p(y_0,y_1,y_2|x),\cY_0,\cY_1,\cY_2)$, and MAC $(\cX_0,\cX_1,\cX_2,q(y|x_0,x_1,x_2),\cY)$.
More specifically, as depicted in Figure~\ref{fig:diamond1}, consider the case in which $\cX=\cY_0=\cY_1=\cX_0=\cX_1=\{0,1\}$; $\cY_2=\cX_2=\{0,1\}^2$; $\cY=\cY_{{\tt W}} \times \cX_2$  for a given memoryless 2-user binary MAC ${\tt W}$: $(\cX_0,\cX_1, p_{{\tt W}}(y_{{\tt W}}|x_0,x_1),\cY_{{\tt W}})$;  for any $x$, $p(Y_0,Y_1,Y_2|x)$ induces $(Y_0,Y_1,Y_2)$ where $Y_0$, $Y_1$, $Z$ are independent Bernoulli(0.5) random variables and $Y_2=(x \oplus Y_Z,Z)$; $X_0^n=f_0(Y_0^n)$, $X_1^n=f_1(Y_1^n)$, $X_2^n=f_2(Y_2^n)$ for relay encoders $f_0$, $f_1$, and $f_2$; and $q(Y|x_0,x_1,x_2)$ for which $Y=(Y_{\tt W},x_2)$. As $Y^n$ holds the value of $X_2^n \in \{0,1\}^n$, which in turn depends on $Y_2^n \in \{0,1\}^n$  through $f_2$, we assume without loss of generality that $X_2^n=Y_2^n$. 

Using the independent nature of relays $Y_0$ and $Y_1$, in   Claim~\ref{claim:3relay} below we tie the cooperation gain of the 2-transmitter MAC {\tt W} with the cooperation gain of the diamond network.

\begin{claim}
\label{claim:3relay}
Let $C_{\text{sum}}(C_{CF})$ be the sum-capacity of the 2-transmitter MAC ${\tt W}$ with user cooperation of rate $C_{CF}$.
Then the capacity $C(C_{CF})$ of the diamond network satisfies $C'(0)=\infty$ if $C'_{\text{sum}}(0)=\infty$.
\end{claim}

\begin{IEEEproof}
See Appendix~\ref{appendix:3relay}.
\end{IEEEproof}

\bibliographystyle{./IEEETran}
\bibliography{diamondrefs}

\clearpage

\appendices

\section{Proof of Theorem~\ref{thm:achievability}}\label{appendix:achievability}

We denote $T_\eps^{(n)}$ as the robustly typical set. See \cite{ElGamal2011} for the definition, as well as the formal statement of the packing lemma, which will be used in the proof.

We employ the following lemma, which is a slight variation on the covering lemma from \cite{ElGamal2011}.

\begin{lemma}\label{lemma:covering}
Let $(U,X,\hatX)\sim p(u,x,\hatx)$ and $\eps'<\eps$. Let $(u^n,x^n)\in T_{\eps'}(U,X)$ be a pair of fixed sequences, and let $\hatX^n(m),m\in\calA$, where $|\calA|\ge 2^{nR}$, be random sequences, conditionally independent of each other, each uniformly distributed on $T_{\eps}^{(n)}(\hatX|u^n)$. Let $M^\star$ be the smallest $m$ for which 
\be
\hatX^n(m)\in T_\eps^{(n)}(\hatX|u^n,x^n).
\ee
If there is no such $m$, we say $M^\star$ is undefined. Then,
\begin{enumerate}
\item there exists $\delta(\eps)$ that tends to zero as $\eps\to 0$ such that $\lim_{n\to\infty} P(M^\star\text{ is undefined})= 0$, if $R>I(X;\hatX|U)+\delta(\eps)$,
\item conditioning on the event that $M^\star$ is defined, $\hatX^n(M^\star)$ is uniformly distributed in $T_\eps^{(n)}(\hatX|u^n,x^n)$.
\end{enumerate}
\end{lemma}
\begin{IEEEproof}
For any $m\in\calA$, we have
\begin{align}
&P((u^n,x^n,\hatX^n(m))\in T_\eps^{(n)}|U^n=u^n,X^n=x^n)
\\&=P((u^n,x^n,\hatX^n(m))\in T_\eps^{(n)}|U^n=u^n)
\\&=\sum_{\hatx^n\in T_\eps^{(n)}(\hatX|u^n,x^n)} \frac{1}{|T_\eps^{(n)}(\hatX|u^n)|}
\\&=\frac{|T_\eps^{(n)}(\hatX|u^n,x^n)|}{|T_\eps^{(n)}(\hatX|u^n)|}
\\&\ge 2^{-n(I(X;\hatX|U)+\delta(\eps))}.
\end{align}
The remainder of the proof of statement 1 follows from an identical argument to that of the standard covering lemma, i.e., \cite[Lemma~3.3]{ElGamal2011}.

To prove statement 2, note first that, if $M^\star$ is defined, then by definition $\hatX^n(M^\star)\in T_\eps^{(n)}(\hatX|u^n,x^n)$. Now, for any $\hatx^n\in T_\eps^{(n)}(\hatX|u^n,x^n)$, 
\begin{align}
&P(\hatX^n(M^\star)=\hatx^n|M^\star\text{ is defined})
\\&=\sum_{m\in \calA} P(M^\star=m|M^\star\text{ is defined}) P\Big(\hatX^n(m)=\hatx^n\Big|\nonumber
\\&\qquad\hatX^n(m')\notin T_\eps^{(n)}(\hatX|u^n,x^n)\text{ for all }m'<m,\nonumber
\\&\qquad \hatX^n(m)\in T_\eps^{(n)}(\hatX|u^n,x^n)\Big)\label{statement2c}
\\&=\sum_{m\in \calA} P(M^\star=m|M^\star\text{ is defined}) P\big(\hatX^n(m)=\hatx^n\big|\nonumber
\\&\qquad\hatX^n(m)\in T_\eps^{(n)}(\hatX|u^n,x^n)\big)\label{statement2d}
\\&=\sum_{m\in \calA} P(M^\star=m|M^\star\text{ is defined}) \frac{1}{|T_\eps^{(n)}(\hatX|u^n,x^n)|}\label{statement2e}
\\&=\frac{1}{|T_\eps^{(n)}(\hatX|u^n,x^n)|}
\end{align}
where \eqref{statement2d} holds since $\hatX^n(m)$ are independent for different $m$, and \eqref{statement2e} since $\hatX^n(m)$ is uniform on $T_\eps^{(n)}(\hatX|u^n)$, and $T_\eps^{(n)}(\hatX|u^n)\subset T_\eps^{(n)}(\hatX|u^n)$. This proves statement 2.
\end{IEEEproof}

We now proceed to the main proof of the theorem. The following argument combines partial-decode-forward/compress-forward strategies, as discussed in Remark~\ref{remark:pdc}.
Fix rates $R_a,R_b,S$ to be determined, where $R_a+R_b=R$. Also fix small constants $0<\eps'<\eps$. We construct a code as follows.

\emph{Codebook generation:}
\begin{itemize}
\item For each $m_a\in[2^{nR_a}]$, generate $u^n(m_a)\sim \prod_{i=1}^n p_U(u_i)$.
\item For each $m_a\in[2^{nR_a}],m_b\in[2^{nR_b}]$, generate $x^n(m_a,m_b)\sim \prod_{i=1}^n p_{X|U}(x_i|u_i(m_a))$.
\item For each $m_a\in[2^{nR_a}]$, $\ell\in [2^{nS}]$, $k\in[2^{nC_{CF}}]$, generate 
$v^n(m_a,\ell,k)\sim \text{Unif}\left[T_\eps^{(n)}(V|u^n(m_a))\right]$, and corresponding source coding bins 
$k\in[2^{nC_{CF}}]$, generate 
$m_0(m_a,\ell,k)\sim\text{Unif}[2^{nC_0}]$.
\item For each $m_a\in[2^{nR_a}]$, $y_r^n\in\calY_r^n$ and $k\in[2^{nC_{CF}}]$, let $\ell(m_a,y_r^n,k)$ be the smallest $\ell\in[2^{nS}]$ such that
\be
(u^n(m_a),y_r^n,v^n(m_a,\ell,k))\in T_\eps^{(n)}(U,Y_r,V).
\ee
If there is no such $\ell$, we say $\ell(m_a,y_r^n,k)$ is undefined.
\end{itemize}

\emph{Encoding:} At the transmitter, given message $m=(m_a,m_b)$, send $x^n(m_a,m_b)$.

\emph{CF coding:} At the CF, given $y_1^n$ and $y_r^n$, first find the unique pair $\hatm_a,\hatm_b$ such that
\be\label{CF_decoding}
(u^n(\hatm_a),x^n(\hatm_a,\hatm_b),y_1^n,y_r^n)\in T_\eps^{(n)}(U,X,Y_1,Y_r).
\ee
Next, find the smallest $k\in[2^{nC_{CF}}]$ such that $\ell(m_a,y_r^n,k)$ is defined, and
\begin{multline}\label{CF_decoding_condition}
(u^n(\hatm_a),x^n(\hatm_a,\hatm_b),y_r^n,y_1^n,v^n(\hatm_a,\ell(\hatm_a,y_r^n,k),k))\\
\in T_\eps^{(n)}(U,X,Y_r,Y_1,V).
\end{multline}
Send this $k$. If there is no such $k$, declare an error.

\emph{Relay coding:} At the relay, given $y_r^n$ and $k$, first find $\hatm_a$ such that
\be
(u^n(\hatm_a),y_r^n)\in T_\eps^{(n)}(U,Y_r).
\ee
Then let $\ell=\ell(\hatm_a,y_r^n,k)$, and send $m_0(\hatm_a,\ell,k)$. If $\ell(\hatm_a,y_r^n,k)$ is undefined, declare an error.

\emph{Decoding:} At the decoder, given $y_1^n$, $m_0$, and $k$, find $\hatm_a,\hatm_b,\hat\ell$ such that
\begin{align}
(u^n(\hatm_a),x^n(\hatm_a,\hatm_b),y_1^n,v^n(\hatm_a,\hat\ell,k))&\in T_\eps^{(n)},\\
m_0(\hatm_a,\hat\ell,k)&=m_0.
\end{align}

\emph{Error analysis:}  Throughout the error analysis, we assume without loss of generality that $m_a=m_b=1$. To prove that $m_a$ is decoded correctly at both the CF and the relay, and that $m_b$ is decoded correctly at the CF, it suffices to consider the following error events:
\begin{align}
\calE_1&=\{(u^n(1),x^n(1,1),y_r^n,y_1^n)\notin T_{\eps'}^{(n)}(U,X,Y_r,Y_1)\},\\
\calE_2&=\{(u^n(m_a),y_r^n)\in T_\eps^{(n)}(U,Y_r)\text{ for some }m_a\ne 1\}\\
\calE_3&=\{(u^n(1),x^n(1,m_b),y_r^n,y_1^n)\in T_\eps^{(n)}(U,X,Y_r,Y_1)\nonumber\\
&\qquad\text{ for some }m_b\ne 1\}.
\end{align}
By the law of large numbers $P(\calE_1)\to 0$. By the packing lemma, $P(\calE_2)\to 0$ and $P(\calE_3)\to 0$ if
\begin{align}
R_a&<I(U;Y_r),\label{Ra_condition}\\
R_b&<I(X;Y_1,Y_r|U).\label{Rb_condition1}
\end{align}

We now show that $\ell(1,y_r^n,k)$ is defined for most values of $k$. For each $k\in [2^{nC_{CF}}]$, define the event
\begin{multline}
\calE_4(k)=\{(u^n(1),y_r^n,v^n(1,\ell,k))\notin T_\eps^{(n)}(U,Y_r,V)
\\ \text{ for all }\ell\in[2^{nS}]\}.
\end{multline}
Since we have already established that $P(\calE_1)\to 0$, by Lemma~\ref{lemma:covering}, $P(\calE_4(k))\to 0$ if
\be\label{S_condition}
S>I(Y_r;V|U).
\ee
Moreover, Lemma~\ref{lemma:covering} asserts that, given $\ell(1,y_r^n,k)$ is defined, $v^n(1,\ell(1,y_r^n,k),k)$ is uniformly distributed on $T_\eps^{(n)}(V|u^n(1),y_r^n)$.
Now consider the event
\be
\calE_5=\left\{|\{k:\ell(1,y_r^n,k)\text{ is defined}\}|<(1-\eps)2^{nC_{CF}}\right\}.
\ee
Since $\ell(1,y_r,k)$ is defined if and only if $\calE_4(k)$ does not occur, it is straightforward to show that $P(\calE_5)\to 0$ as $n\to\infty$. Now consider the error event in which the CF cannot find a value of $k$ to transmit, i.e.,
\begin{multline}
\!\calE_6=\{(u^n(1),x^n(1,1),y_r^n,y_1^n,v^n(1,\ell(1,y_r^n,k),k))\notin T_\eps^{(n)}\\
\text{ for all }k\in[2^{nC_{CF}}]\}.
\end{multline}
We may now apply Lemma~\ref{lemma:covering} a second time to find that $P(\calE_6)\to 0$ if
\be
C_{CF}>I(X,Y_1;V|U,Y_r).
\ee

Let us further assume without loss of generality that $k=1$, $\ell(1,y^n,1)=1$, and $m_0(1,1,1)=1$. Assuming that error events $\calE_1,\calE_2,\calE_3,\calE_6$ do not occur, the relay  selects $\hatm_a=\ell=m_0=1$, and
\begin{align}
(u^n(1),x^n(1,1),y_r^n,y_1^n,v^n(1,1,1))&\in T_\eps^{(n)}.
\end{align}
Now consider the following decoding error events:
\begin{align}
\calE_7&=\{(u^n(1),x^n(1,m_b),y_1^n,v^n(1,1,1))\in T_\eps^{(n)}\nonumber
\\&\qquad\text{ for some }m_b\ne 1\},\\
\calE_8&=\{(u^n(1),x^n(1,m_b),y_1^n,v^n(1,\ell,1))\in T_\eps^{(n)},\nonumber
\\&\qquad m_0(1,\ell,1)=1\text{ for some }m_b\ne 1,\ell\ne 1\},\\
\calE_9&=\{(u^n(m_a),x^n(m_a,m_b),y_1^n,v^n(m_a,\ell,1))\in T_\eps^{(n)},\nonumber
\\&\qquad m_0(m_a,\ell,1)=1\text{ for some }m_a\ne 1,m_b,\ell\}.
\end{align}
Applying the packing lemma several times, $P(\calE_7)\to 0$ if
\be\label{Rb_condition2}
R_b<I(X;Y_1,V|U),
\ee
$P(\calE_8)\to 0$ if
\be\label{Rb_condition3}
R_b+S<I(X;Y_1|U)+I(V;X,Y_1|U)+C_0,
\ee
and $P(\calE_9)\to 0$ if
\be\label{R_condition}
R_a+R_b+S<I(U,X;Y_1)+I(V;X,Y_1|U)+C_0.
\ee

We now collect the various rate conditions required for all of the error event probabilities to vanish. It is advantageous if $S$ is as small as possible; from the lower limit in \eqref{S_condition}, we may assume that $S$ is slightly larger than $I(Y_r;V|U)$. We have three conditions on $R_b$, namely \eqref{Rb_condition1}, \eqref{Rb_condition2}, and \eqref{Rb_condition3}. Combining each of these with the condition on $R_a$ in \eqref{Ra_condition}, and recalling that $R=R_a+R_b$, we need
\begin{align}
R&<I(U;Y_r)+I(X;Y_1,Y_r|U),\\
R&<I(U;Y_r)+I(X;Y_1,V|U),\\
R&<I(U;Y_r)+I(X;Y_1|U)+I(V;X,Y_1|U)+C_0\nonumber\\
&\qquad-I(Y_r;V|U).
\end{align}
Furthermore, from \eqref{R_condition} we need
\begin{align}
R&<I(U,X;Y_1)+I(V;X,Y_1|U)-I(Y_r;V|U)+C_0
\\&=I(U;Y_1)+I(X;Y_1|U)+I(V;X,Y_1|U)\nonumber\\
&\qquad-I(Y_r;V|U)+C_0.
\end{align}
Therefore, the conditions in the statement of the theorem imply that $R_a,R_b,S$ can be found such that all of the above conditions hold.

\section{Proof of Theorem~\ref{thm:infinite_slope}}\label{appendix:infinite_slope}

Under the starting distribution
\be
p(u,x)p(y_1,y_r|x)p(v|u,y_r),
\ee
$I(X,Y_1;V|U,Y_r)=0$. To show \eqref{infinite_slope}, we modify this distribution slightly, in a way that gives $I(X,Y_1;V|U,Y_r)>0$, which corresponds to positive $C_{CF}$, while increasing the achieved rate. In particular, we leave $p(u,x)$ fixed, but change the conditional distribution for $v$ to
\be
q(v|u,x,y_1,y_r)=p(v|u,y_r)+\alpha\, r(v|u,x,y_1,y_r)
\ee 
where $\alpha\approx 0$. For a variable $A\subset\{U,X,Y_1,Y_r\}$, we further define $r(v|a)$, for example by
\be
r(v|u,y_r)=\sum_{x,y_1} p(x,y_1|u,y_r)r(v|u,x,y_1,y_r).
\ee
Thus $q(v|a)=p(v|a)+\alpha\, r(v|a)$. In order for $q$ to be a valid distribution, we need
\be
\sum_v r(v|u,x,y_1,y_r)=0\text{ for all }u,x,y_1,y_r.
\ee
Thus, these $r$ functions are not really distributions; instead they satisfy $\sum_v r(v|a)=0$ for any variable $A$. Moreover, if $p(v|u,y_r)=0$, then in order for $q$ to be valid, we need $r(v|u,x,y_1,y_r)\ge 0$; we here make the simplifying assumption that $r(v|u,x,y_1,y_r)=0$ for any $u,y_r,v$ where $p(v|u,y_r)=0$. This assumption has the following consequence. Suppose for some $(u,x,y_1,y_r,v)$, $p(u,x,y_1,y_r,v)=0$. Recalling 
\be
p(u,x,y_1,y_r,v)=p(u,x,y_1,y_r) p(v|u,y_r)
\ee
it must be true that either $p(u,x,y_1,y_r)$ or $p(v|u,y_r)$ is zero. That is, if $p(u,x,y_1,y_r)>0$, then $r(v|u,x,y_1,y_r)=0$, so $q(v|u,y_1,y_r,u)=0$. In particular
\be
p(u,x,y_1,y_r)q(v|u,x,y_1,y_r)
\ee
vanishes if and only if $p(u,x,y_1,y_r,v)$ vanishes.

We are changing the distribution of $V$, but not of $U$, so many terms in the lower bounds in Thm.~\ref{thm:achievability} do not change with $\alpha$. Define the following functions:
\begin{align*}
f_1(\alpha)&=I_q(X;V|U,Y_1),
\\
f_2(\alpha)&=I_q (V;X,Y_1|U)-I_q(Y_r;V|U),
\\
C_{CF}(\alpha)&=I_q(X,Y_1;V|U,Y_r).
\end{align*}
Note that $f_1$ and $f_2$ include all the terms that change with $V$ in \eqref{rate_condition1} and \eqref{rate_condition2} respectively. Since by assumption, $I(X;Y_1,V|U)<I(X;Y_1,Y_r|U)$, if $f_1$ increases with $\alpha$ then so does the right-hand side of \eqref{rate_condition1}. Thus, to prove \eqref{infinite_slope}, it is enough for
\be
C_{CF}'(0)=0,\quad f_1'(0)>0,\quad f_2'(0)>0.
\ee

We first show that, given the assumptions we have already made, $C_{CF}'(0)=0$. We have
\begin{align}
&C_{CF}(\alpha)=I_q(X,Y_1;V|U,Y_r)
\\&=D(q(u,x,y_1,y_r,v)\|q(u,x,y_1,y_r)q(v|u,y_r))
\\&=\sum_{u,x,y_1,y_r}p(u,x,y_1,y_r) D(q(v|u,x,y_1,y_r)\|q(v|u,y_r)).\label{C_CF}
\end{align}
Note that $p(v|x,y_1,y_r,u)=p(v|u,y_r)$, so $C_{CF}(0)=0$. To find $C_{CF}'(\alpha)$, consider an arbitrary function of the form
\begin{align}
f(\alpha)&=D(p(x)+\alpha r_1(x)\|p(x)+\alpha r_2(x))
\\&=\sum_x (p(x)+\alpha r_1(x))\log \frac{p(x)+\alpha r_1(x)}{p(x)+\alpha r_2(x)}
\end{align}
where $\sum_x r_i(x)=0$ for $i=1,2$, and $r_1(x)=r_2(x)=0$ whenever $p(x)=0$. Thus, the only relevant terms in the summation are where $p(x)>0$, so
\begin{align}
f'(0)&=\lim_{\alpha\to 0} f'(\alpha)
\\&=\lim_{\alpha\to 0} \sum_{x:p(x)>0} \bigg[r_1(x)\log \frac{p(x)+\alpha r_1(x)}{p(x)+\alpha r_2(x)}
+r_1(x)\\
&\qquad-r_2(x)\frac{p(x)+\alpha r_1(x)}{p(x)+\alpha r_2(x)}\bigg]
\\&=\sum_{x:p(x)>0} (r_1(x)-r_2(x))
\\&=0.
\end{align}
To apply this analysis to the function $C_{CF}(\alpha)$ from \eqref{C_CF}, given any $u,x,y_1,y_r$, consider
\be
D(q(v|u,x,y_1,y_r)\|q(v|u,y_r)).
\ee
Recall that
\begin{align}
q(v|u,x,y_1,y_r)&=p(v|u,y_r)+\alpha\, r(v|u,x,y_1,y_r),\\
q(v|u,y_r)&=p(v|u,y_r)+\alpha\, r(v|u,y_r).
\end{align}
Moreover, we have made the assumption that $r(v|u,x,y_1,y_r)=0$ whenever $p(v|u,y_r)=0$, so we have a scenario matching the above assumptions on $f(\alpha)$. Thus, $C_{CF}'(0)=0$.

We now consider the conditions when $f_1'(0),f_2'(0)>0$. Consider a variable $A\subset\{X,Y_1,Y_r,U\}$. Recalling the fact that if $p(a)>0$ and $p(v|a)=0$, then $q(v|a)=0$, we may write
\begin{align}
&H_q(V|A)=-\sum_{a,v:q(v|a)>0} p(a) q(v|a)\log q(v|a)
\\&=-\sum_{\substack{u,x,y_1,y_r,v:\\
p(u,x,y_1,y_r)>0,\\  q(v|u,x,y_1,y_r)>0}} p(u,x,y_1,y_r)q(v|u,x,y_1,y_r)\log q(v|a)
\\&=-\sum_{\substack{u,x,y_1,y_r,v:\\
p(u,x,y_1,y_r,v)>0}}p(u,x,y_1,y_r)q(v|u,x,y_1,y_r)\log q(v|a)
\end{align}
where we have used the fact that $p$ and $q$ have precisely the same support.
Thus
\begin{align}
&\frac{d}{d\alpha} H_q(V|A)
\\&=-\!\!\!\!\!\!\sum_{\substack{u,x,y_1,y_r,v:\\
p(u,x,y_1,y_r,v)>0}} p(u,x,y_1,y_r) \frac{d}{d\alpha} q(v|u,x,y_1,y_r)\log q(v|a)
\\&=-\sum_{\substack{u,x,y_1,y_r,v:\\
p(u,x,y_1,y_r,v)>0}} p(u,x,y_1,y_r) \bigg[r(v|u,x,y_1,y_r)\log q(v|a)\nonumber
\\&\qquad+\frac{q(v|u,x,y_1,y_r)r(v|a)}{q(v|a)}\bigg]
\\&=-\sum_{\substack{u,x,y_1,y_r,v:\\p(u,x,y_1,y_r,v)>0}}
p(u,x,y_1,y_r)r(v|u,x,y_1,y_r)\log q(v|a)\nonumber
\\&\qquad+\sum_{a,v} p(a)r(v|a)
\\&=-\sum_{\substack{u,x,y_1,y_r,v:\\p(u,x,y_1,y_r,v)>0}}
p(u,x,y_1,y_r)r(v|u,x,y_1,y_r)\log q(v|a).
\end{align}
In particular,
\begin{multline}
\frac{d}{d\alpha} H_q(V|A)\bigg|_{\alpha=0}
\\=-\!\!\!\!\!\!\sum_{\substack{u,x,y_1,y_r,v:\\p(u,x,y_1,y_r,v)>0}}
p(u,x,y_1,y_r)r(v|u,x,y_1,y_r)\log p(v|a).
\end{multline}

Now we may easily write
\begin{align}
f_1'(0)&=\sum_{\substack{u,x,y_1,y_r,v:\\p(u,x,y_1,y_r,v)>0}}
p(u,x,y_1,y_r)r(v|u,x,y_1,y_r)\nonumber
\\&\qquad\cdot\log \frac{p(v|u,x,y_1)}{p(v|u,y_1)},\\
f_2'(0)&=\sum_{\substack{u,x,y_1,y_r,v:\\p(u,x,y_1,y_r,v)>0}}
p(u,x,y_1,y_r) r(v|u,x,y_1,y_r)\nonumber
\\&\qquad\cdot \log \frac{p(v|u,x,y_1)}{p(v|u,y_r)}.
\end{align}

Recall that we are interested in showing that $f_1'(0),f_2'(0)>0$. Since each of these is a linear function of $r$, we consider a generic linear set up. In particular, we are interested in whether there exists a vector $z$ such that $a^Tz>0$, $b^Tz>0$, and $Az=0$. 
That is, we are interested in
\begin{align}
&\max_{z:Az=0} \min\{a^Tz,b^Tz\}
\\&=\max_{z:Az=0} \min_{\lambda\in[0,1]} \lambda a^Tz+(1-\lambda) b^Tz
\\&=\max_{z} \min_{\lambda\in[0,1],\gamma}  \lambda a^Tz+(1-\lambda) b^Tz+\gamma^T Az
\\&=\max_z  \min_{\lambda\in[0,1],\gamma} (\lambda a+(1-\lambda)b+A^T\gamma)^Tz
\\&=\min_{\lambda\in[0,1],\gamma} \max_z  (\lambda a+(1-\lambda)b+A^T\gamma)^Tz
\\&=\min_{\lambda\in[0,1],\gamma} \begin{cases}0 & \lambda a+(1-\lambda)b+A^T\gamma=0\\ \infty & \text{otherwise}.\end{cases}
\end{align}
That is, there exists no $z$ of interest if and only if there exists $\lambda\in[0,1]$ and $\gamma$ where
\be
\lambda a+(1-\lambda)b+A^T\gamma=0.
\ee

Applying this principle to our situation, there does \emph{not} exist such an $r$ function if and only if there exists $\lambda\in[0,1]$, $\gamma(u,x,y_1,y_r)$ where
\begin{multline}
p(u,x,y_1,y_r) \bigg[\lambda \log \frac{p(v|u,x,y_1)}{p(v|y_1)}
\\+(1-\lambda)\log \frac{p(v|u,x,y_1)}{p(v|u,y_r)}\bigg]
+\gamma(u,x,y_1,y_r)=0,\\
\text{for all }x,y_1,y_r,v:p(u,x,y_1,y_r,v)>0.
\end{multline}
Dividing by $p(u,x,y_1,y_r)$  and rearranging gives
\begin{multline}
p(v|u,x,y_1)=\frac{p(v|u,y_1)^\lambda p(v|u,y_r)^{1-\lambda}}{\gamma(u,x,y_1,y_r)}
\\
\text{ for all }x,y_1,y_r,v:p(u,x,y_1,y_r,v)>0.
\end{multline}

\section{Proof of Corollary~\ref{corollary:all_positive_channel}}\label{appendix:all_positive_channel}

It is enough to show that either Possibility~1 in the corollary statement holds, or the sufficient condition in Thm.~\ref{thm:infinite_slope} holds. Thus, it is enough to prove that, if the sufficient condition in Thm.~\ref{thm:infinite_slope} does \emph{not} hold, then Possibility~1 must hold. That is, we assume \eqref{infinite_slope_condition} holds for all $u,x,y_1,y_r,v$ such that $p(u,x,y_1,y_r)>0,p(v|y_r,u)>0$, and we prove the existence of a function $g$. We may assume without loss of generality that $U$ has full support, since if not we may simply delete any zero-probability letters. Fix any $u\in\calU$, and let $x\in\calX_u$, where 
\be
\calX_u=\{x\in\calX:p(x|u)>0\}.
\ee
Thus, by the assumption of the corollary, for any $y_1,y_r$,
\be
p(u,x,y_1,y_r)=p(u,x)p(y_1,y_r|x)>0.
\ee
In the following portion of the proof, we will only focus on this $u$ value, and so for convenience we will drop the dependence on $u$ in the conditional distributions. That is, we re-write \eqref{infinite_slope_condition} as
\be\label{infinite_slope_condition_no_u}
p(v|x,y_1)=\frac{p(v|y_1)^\lambda p(v|y_r)^{1-\lambda}}{\gamma(x,y_1,y_r)}
\ee
which must hold for all $x,y_1,y_r,v$ where $x\in\calX_u$ and $p(v|y_r)>0$.

Consider the graph on vertex set $\calV$ with edge set given by
\begin{multline}
\calE=\{(v_a,v_b):\text{ there exists }y_r\in\calY_r\text{ such that}\\ p(v_a|y_r),p(v_b|y_r)>0\}.
\end{multline}
Consider any pair $(v_a,v_b)\in\calE$. By definition there exists a $y_r$ such that  $p(v_a|y_r),p(v_b|y_r)>0$. For any $x\in\calX_u,y_1\in\calY_1$, from \eqref{infinite_slope_condition_no_u} we have
\begin{align}
\frac{p(v_a|x,y_1)}{p(v_b|x,y_1)}&=
\frac{p(v_a|y_1)^\lambda p(v_a|y_r)^{1-\lambda}}{\gamma(x,y_1,y_r)}
\\&\qquad
\cdot\frac{\gamma(x,y_1,y_r)}{p(v_b|y_1)^\lambda p(v_b|y_r)^{1-\lambda}}
\\&=
\left(\frac{p(v_a|y_1)}{p(v_b|y_1)}\right)^\lambda \left(\frac{p(v_a|y_r)}{p(v_b|y_r)}\right)^{1-\lambda}.\label{beta_quantity}
\end{align}
We define the quantity in \eqref{beta_quantity} as $\beta(v_a,v_b,y_1)$. If there is more than one valid $y_r$, then by \eqref{beta_quantity} they each much produce the same value. Thus for all $x\in\calX_u$ and $y_1$,
\be
\frac{p(v_a|x,y_1)}{p(v_b|x,y_1)}=\beta(v_a,v_b,y_1).
\ee
Suppose $v_a$ and $v_b$ are connected in the graph $(\calV,\calE)$. That is, there exists a sequence of letters $v_a=v_1,\ldots,v_k=v_b$ where
\be
(v_1,v_2),\ldots,(v_{k-1},v_k)\in\calE.
\ee
Thus, for all $x,y_1$,
\be
\frac{p(v_a|x,y_1)}{p(v_b|x,y_1)}=\prod_{\ell=1}^{k-1} \frac{p(v_\ell|x,y_1)}{p(v_{\ell+1}|x,y_1)}
=\prod_{\ell=1}^{k-1} \beta(v_\ell,v_{\ell+1},y_1).
\ee
We may define the latter as $\beta(v_a,v_b,y_1)$ for any connected $v_a,v_b$.

The graph $(\calV,\calE)$ splits into connected sub-graphs with vertex sets $\calV_1,\calV_2,\ldots,\calV_m$, where these vertex sets represent a partition of $\calV$. Define a random variable $W$, with alphabet $\{1,\ldots,m\}$, where $W=w$ whenever $V\in\calV_w$. Thus $W$ is a deterministic function of $V$. Moreover, $W$ is a deterministic function of $Y_r$, since for any $y_r$, all letters $v$ where $p(v|y_r)>0$ must be in the same sub-graph $\calV_w$ for some $w$. Consider any $w\in\{1,\ldots,m\}$. Let $v_w$ be a designated element of $\calV_w$. For any $v\in\calV_w$, we have
\be
\frac{p(v|x,y_1)}{p(v_w|x,y_1)}=\beta(v,v_w,y_1)\text{ for all }x.
\ee
Thus, for any $v\in\calV_w$
\begin{align}
p(v|x,y_1,w)
&=\frac{p(v,w|x,y_1)}{p(w|x,y_1)}
\\&=\frac{p(v|x,y_1)}{p(w|x,y_1)}
\\&=\frac{p(v|x,y_1)}{\sum_{v'\in\calV_w} p(v'|x,y_1)}
\\&=\frac{p(v|x,y_1)/p(v_w|x,y_1)}{\sum_{v'\in\calV_w} p(v'|x,y_1)/p(v_w|x,y_1)}
\\&=\frac{\beta(v,v_w,y_1)}{\sum_{v'\in\calV_w} \beta(v',v_w,y_1)}.
\end{align}
If $v\notin\calV_w$, then obviously $p(v|x,y_1,w)=0$. Thus
\be
p(v|x,y_1,w)=1(v\in\calV_w) \frac{\beta(v,v_w,y_1)}{\sum_{v'\in\calV_w} \beta(v',v_w,y_1)}.
\ee

Since the RHS does not depend on $x$, we must have
\be
p(v|x,y_1,w)=p(v|y_1,w).
\ee

We now reintroduce the dependence on $u$. Since the above analysis holds for any $u\in\calU$, it must be that
\be
p(v|u,x,y_1,w)=p(v|u,y_1,w)
\ee
That is, $X-(U,Y_1,W)-V$ is a Markov chain. Recall that by assumption, rate $R$ satisfies \eqref{pdc_bd1}--\eqref{pdc_bd2}. We show that replacing $V$ by $W$ does not reduce the achieved rate in this bound. For \eqref{pdc_bd1}, note that
\begin{align}
I(X;Y_1,V|U)&=I(X;Y_1,W,V|U)
\\&=I(X;Y_1,W|U)+I(X;V|U,Y_1,W)
\\&=I(X;Y_1,W|U).
\end{align}
For \eqref{pdc_bd2}, note that
\begin{align}
    I(Y_r;V|U,X,Y_1)&=I(Y_r;W,V|U,X,Y_1)
    \\&\ge I(Y_r;W|U,X,Y_1).
\end{align}
Thus, $R$ must satisfy \eqref{pdc_bd1}--\eqref{pdc_bd2} with $V$ replaced by $W$. The proof is completed by recalling that $W$ is a deterministic function of $U$ and $Y_r$.

\section{Proof of Claim~\ref{claim:3relay}}\label{appendix:3relay}

We first show that $C(0) \leq C_{sum}(0)/2$.
We then show that $C(C_{CF}) \geq C_{sum}(C_{CF})/2$ for any $C_{CF}>0$.
Together, these imply our assertion.

Let $\varepsilon >0$ be an arbitrarily small parameter, and let $n$ be sufficiently large.
To show that $C(0) \leq C_{sum}(0)/2$, consider any functions $f_0^n(Y_0^n)=X_0^n$ and $f_1^n(Y_1^n)=X_1^n$. For $n$ sufficiently large, 
\begin{align*}
C(0)n-\varepsilon n  \leq & I(X^n;Y^n) \\
= & I(X^n;Y_2^n)+ I(X^n;Y^n_{\tt W}|Y_2^n) \\
= & I(X^n;Y^n_{\tt W}|Y_2^n) \\
= & I(X^n;Y^n_{\tt W}|X^n \oplus Y_{Z^n}^n,Z^n) \\
\leq  & I(X^n, X^n \oplus Y_{Z^n}^n;Y^n_{\tt W}|Z^n)\\
=  & I(X^n, Y_{Z^n}^n;Y^n_{\tt W}|Z^n)\\
=  & I(Y_{Z^n}^n;Y^n_{\tt W}|Z^n) +I(X^n;Y^n_{\tt W}|Y_{Z^n}^n,Z^n) \\ 
=  & I(Y_{Z^n}^n;Y^n_{\tt W}|Z^n)\\
\stackrel{(a)}{=}  & \frac{1}{2}(I(Y_{Z^n}^n;Y^n_{\tt W}|Z^n) + I(Y_{(Z^n)^c}^n;Y^n_{\tt W}|Z^n)) \\
\leq  & \frac{1}{2}(I(Y_{Z^n}^n;Y^n_{\tt W}|Z^n) + I(Y_{(Z^n)^c}^n;Y^n_{\tt W},Y_{Z^n}^n|Z^n))\\
=  & \frac{1}{2}(I(Y_{Z^n}^n;Y^n_{\tt W}|Z^n) +I(Y_{(Z^n)^c}^n;Y_{Z^n}^n|Z^n) \\
& \hspace{1.2cm} +I(Y_{(Z^n)^c}^n;Y^n_{\tt W}|Y_{Z^n}^n,Z^n)) \\
=  & \frac{1}{2}(I(Y_{Z^n}^n;Y^n_{\tt W}|Z^n) +I(Y_{(Z^n)^c}^n;Y^n_{\tt W}|Y_{Z^n}^n,Z^n))\\
=  & \frac{1}{2}I(Y_{Z^n}^n,Y_{(Z^n)^c}^n;Y^n_{\tt W}|Z^n) \\
=  & \frac{1}{2}I(Y_{0}^n,Y_{1}^n;Y^n_{\tt W}) \leq \frac{C_{sum}(0)}{2}n.
\end{align*}
In ({\em a}) above we use the equality $I(Y_{Z^n}^n;Y^n_{\tt W}|Z^n) = I(Y_{(Z^n)^c}^n;Y^n_{\tt W}|Z^n))$ which follows from the independent and symmetric nature of $Z^n$.
Thus, the capacity of the diamond network above is at most half the sum-capacity of ${\tt W}$.

To show that $C(C_{CF}) \geq C_{sum}(C_{CF})/2$, consider any rate vector $(r_0,r_1)$ achievable on the 2-transmitter MAC ${\tt W}$ with a cooperation facilitator of rate $C_{CF}$. Assume that $r_0\geq r_1$ (a symmetric argument is used otherwise). We construct a rate $r=(r_0+r_1)/2$ scheme for the diamond network in which the first two relays cooperate at rate $C_{CF}$. Consider any $rn$ bit message $M$ for the diamond network. We treat $M$ as two messages, an $r_1 n$-bit message $M_1$ and an  $(rn-r_1n)$-bit message $M_2$ with $rn-r_1n=(r_0-r_1)n/2$. Encode $M$ to a binary word $X^n$ using a three-part code: the first $r_1 n$ bits of $X^n$ equal $M_1$, the next $(r_0-r_1)n$ bits of $X^n$ are a rate-$(1/2)$ erasure encoding of $M_2$, the remaining bits of $X^n$ are all set to zero.
Let message $m_0$ for ${\tt W}$ be the first $r_0n$ bits of $Y_0^n$. Let message $m_1$ for ${\tt W}$ be the first $r_1n$ bits of $Y_1^n$. Let $X_0^n=f_0^n(Y_0^{r_0n})$ and $X_1^n=f_1^n(Y_1^{r_1n})$, where $f_1^n$ and $f_2^n$ are the encoding functions of ${\tt W}$ that achieve rate vector $(r_0,r_1)$. 
Now, from the outcome $Y^n_{\tt W}$ of ${\tt W}$, the first $r_0n$ bits of $Y_0^n$ and the first $r_1n$ bits of $Y_1^n$ can be decoded (using the decoder of ${\tt W}$). This implies, using $Y_2^n$, that the first $r_1n$ bits of $X^n$ (and thus $M_1$) can be decoded.
Moreover, roughly speaking, out of the next $(r_0-r_1)n$ bits of $Y_0^n$ (approximately) $(r_0-r_1)n/2$ bits of $X^n$ (at random locations according to $Z^n$) can be decoded, implying that $M_2$ can also be decoded using the capacity 1/2 erasure decoder. All in all, $M$ is decoded successfully.
To be more precise, for any $\varepsilon>0$, in the arguments above one defines $r$ to be $\frac{r_0+r_1}{2}-\varepsilon$, uses a rate $\frac{1}{2}-\varepsilon$ erasure code, and, through standard concentration, shows that indeed a rate-$r$ message $M$ is decoded successfully with probability that depends on $\varepsilon$ and tends to one when $\varepsilon$ tends to zero. As $\varepsilon$ is arbitrary, we conclude that $C(C_{CF}) \geq C_{sum}(C_{CF})/2$.

\end{document}